\documentclass[10pt,twocolumn]{article}
\usepackage[english]{babel}
\usepackage[activate={true,nocompatibility},final,tracking=true,kerning=true,spacing=,factor=1100,stretch=10,shrink=10]{microtype}
\usepackage{authblk} 
\usepackage[separate-uncertainty=true]{siunitx}
\usepackage{amsmath,bm}
\usepackage{graphicx}
\usepackage[nomove,nospace]{cite}
\usepackage{float}
\usepackage[all]{nowidow}
\usepackage[font=small]{caption}
\usepackage[left=48pt, right=42pt,top=46pt,bottom=60pt,headheight=15pt,headsep=10pt,letterpaper,twoside]{geometry}
\usepackage{titlesec}
\usepackage{lineno}
\usepackage{hhline}
\usepackage{lipsum} 
\usepackage{caption}
\usepackage{mathtools}
\usepackage[shortcuts]{extdash} 
\usepackage[dvipsnames]{xcolor}
\usepackage[colorlinks=true,linkcolor=black,citecolor=black,urlcolor=black]{hyperref} 
\author[1,2]{Johannes~Bütow}
\author[1,3]{Varun~Sharma}
\author[1]{Dorian~Brandmüller}
\author[1,2,4,5]{Jörg~S.~Eismann}
\author[1,2,3,4,5,$^*$]{Peter~Banzer}
\affil[1]{Institute of Physics, University of Graz, NAWI Graz, Universitätsplatz 5, 8010 Graz, Austria}
\affil[2]{Christian Doppler Laboratory for Structured Matter Based Sensing, Institute of Physics, Universitätsplatz 5, 8010 Graz, Austria}
\affil[3]{Max\,Planck-University\,of\,Ottawa\,Centre\,for\,Extreme\,and\,Quantum\,Photonics, 25\,Templeton\,St., Ottawa, Ontario\,K1N\,6N5, Canada}
\affil[4]{Max Planck Institute for the Science of Light, Staudtstr. 2, 91058 Erlangen, Germany}
\affil[5]{Institute of Optics, Information and Photonics, University Erlangen-Nuremberg, Staudtstr. 7/B2, 91058 Erlangen, Germany}
\affil[*]{peter.banzer@uni-graz.at}
\date{} 
\DeclareCaptionLabelSeparator{bar}{\space\textbar\space}

\titleformat{\section}{\centering\normalfont\fontsize{12}{15}\bfseries}{\thesection}{0.5em}{}
\title{Photonic integrated processor for structured light detection and distinction} 
\makeatletter 
\renewcommand*{\@biblabel}[1]{\hfill#1.} 
\makeatother

\begin{document}
\maketitle

\begin{bfseries}
\noindent
Integrated photonic devices have become pivotal elements across most research fields that involve light-based applications.
A particularly versatile category of this technology are programmable photonic integrated processors, which are being employed in an increasing variety of applications, like communication or photonic computing. 
Such processors accurately control on-chip light within meshes of programmable optical gates.
Free-space optics applications can utilize this technology by using appropriate on-chip interfaces to couple distributions of light to the photonic chip.
This enables, for example, access to the spatial properties of free-space light, particularly to phase distributions, which is usually challenging and requires either specialized devices or additional components.
Here we discuss and show the detection of amplitude and phase of structured higher-order light beams using a multipurpose photonic processor.
Our device provides measurements of amplitude and phase distributions which can be used to, e.g., directly distinguish light's orbital angular momentum without the need for further elements interacting with the free-space light.
Paving a way towards more convenient and intuitive phase measurements of structured light, we envision applications in a wide range of fields, specifically in microscopy or communications where the spatial distributions of lights properties are important.
\end{bfseries}

\section*{Introduction}

\begin{figure*}[thb] 
  \centering\includegraphics[width = 1\linewidth]{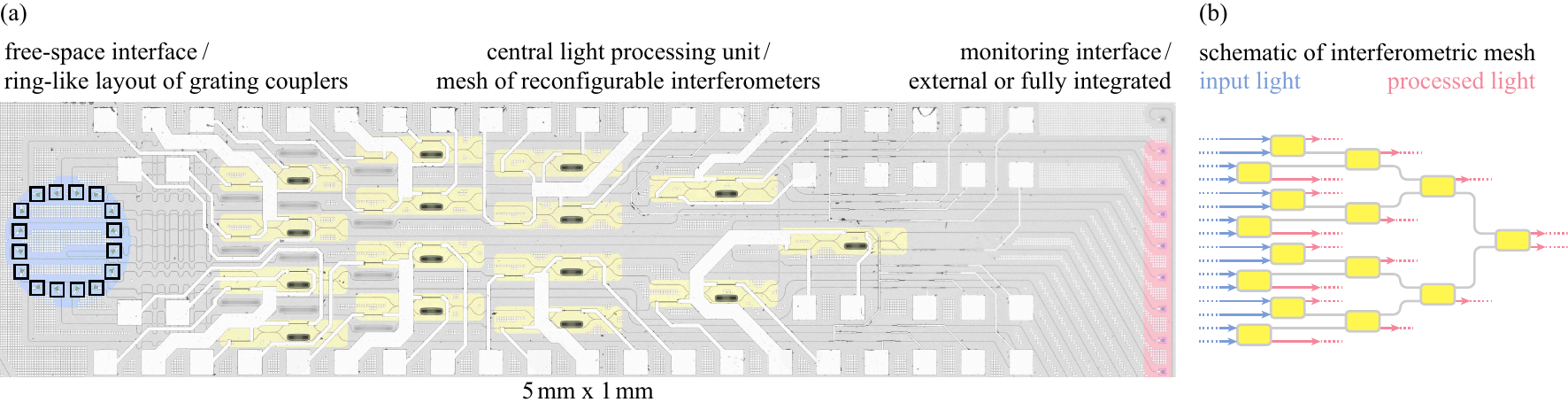}\caption{
    (a)~Optical microscope image of the multipurpose photonic integrated processor.
    On-chip light flows from left-to-right and is processed losslessly in a central mesh of reconfigurable Mach-Zehnder interferometers.
    (b)~Schematic view of the central light processing unit's waveguide and interferometer layout.
    }\label{fig:ChipImage}
\end{figure*}

The small-footprint integration of optical and photonic components allows for on-chip processing via controlled routing, interaction, and manipulation of light. 
Such photonic integrated circuits have paved the pathway towards the development of novel devices across various research areas \cite{Chen.2018b}.
More recently, actively controlled reconfigurable photonic integrated circuits have emerged and are being explored experimentally \cite{Harris.2018, Bogaerts.2020b, Bogaerts.2020}.
These chips process waveguide modes inside a reprogrammable light processing unit with the help of meshes of optical gates\,/\,Mach-Zehnder interferometers and thus losslessly manipulate relative on-chip amplitudes and phases across the photonic circuit.
Tailoring, rerouting and assessing on-chip light in such interferometric meshes has enabled applications across various fields, like communication \cite{Annoni.2017}, information processing and quantum optics \cite{Carolan.2015, Perez.2017} and photonic computing and neural networks \cite{Shen.2017,Pai.2023b,Pai.2023}.  
Programmable photonic processors like these can also be interfaced to free-space, e.g., via grating coupler based layouts.
Operating where on-chip waveguide modes meet off-chip free-space light, the resulting photonic chips can be utilized in different ways and used for multiple purposes in free-space optics \cite{Miller.2013b,Miller.2013}.  
This can, for example, allow the targeted generation of structured light \cite{Butow.2023}, the measurement of amplitude and phase \cite{Miller.2020,Butow.2022}, the separation of arbitrary modes \cite{Milanizadeh.2022} or the coherent self-control of free-space modes \cite{Milanizadeh.2021}.

Such a free-space interface intricately connects these devices to distributions of free-space modes and their natural next application is the measurement and distinction of higher-order light beams that feature intricate amplitude and phase distributions.
Such beams are facilitated in advanced applications like microscopy, communication or quantum entanglement \cite{Mair.2001, Hell.2007, Willner.2015, RubinszteinDunlop.2017, Willner.2021}.
However, intensity-only measurements are usually blind to phase or polarization of such beams, although information is usually encoded in these parameters.
For their characterization, more elaborate approaches are required.
For example, identifying the helical phase-front of a Laguerre-Gaussian beam and the related property of orbital angular momentum \cite{Allen.1992} requires phase sensitive spatially or angularly resolved measurements, which can be experimentally challenging.
Consequently, considerable research has been focusing on the convenient identification of such beams with orbital angular momentum \cite{Fatkhiev.2021} using interferometry, Shack-Hartmann wavefront sensors, spatial light modulators \cite{Berkhout.2010,Forbes.2016} or specifically designed integrated components \cite{Rui.2016, Chen.2018}.

Here we discuss and demonstrate the detection and distinction of structured higher-order light beams, including but not limited to Laguerre-Gaussian modes, using a photonic integrated processor.
We specifically show how such a device, even with only 16 pixels, can distinguish a wide range of beams, e.g., Laguerre-Gaussian modes carrying orbital angular momentum. 
This is an important step towards an all-integrated, versatile and high-resolution platform for the measurement and control of amplitude and phase distributions of such beams, going far beyond information and functionality offered by a conventional camera.
The results presented in this article extend the range of applications of multipurpose photonic processors, highlighting their strength in terms of controlling on-chip light for free-space applications.

\section*{Main} 
\noindent
\textbf{On-chip processing of light.}
The photonic integrated processor is based on a \SI{220}{\nano\meter} silicon-on-insulator platform.
A die shot of the chip is shown in Fig.\,\ref{fig:ChipImage}\,(a).
The utilized processor is composed of three major sections, each including multiple integrated components: 1)~A central light processing unit comprised of a mesh of reconfigurable Mach-Zehnder interferometers (universal 2$\times$2 optical gates), highlighted in yellow and also shown schematically in Fig.\,\ref{fig:ChipImage}\,(b).
2)~A free-space interface connected to the mesh via single-mode waveguides, highlighted in blue, and 3)~a monitoring interface, highlighted in red.
The free-space and monitoring interface are composed of carefully arranged grating couplers terminating the individual waveguides.
These grating couplers can either act as emitters to free space or they can be used to couple free-space light into the waveguides, sampling impinging light beams comparable to the pixels of a normal camera.
Beyond the capabilities of conventional pixels, however, the phase information of sampled light is preserved for the integrated photonic processor in the coupled waveguide modes that now travel across the chip. 
This is essential for the measurement of structured free-space light discussed in this manuscript.
In reverse, when grating couplers couple out on-chip light into free space, the emitted light has the same relative intensities and phases as the terminated waveguide modes.
Grating couplers can thus be utilized for off-chip power monitoring of waveguide modes transmitted through the interferometric mesh by means of imaging or fiber coupling via the monitoring interface (red).
Alternatively, power monitoring could be fully integrated, thus no longer requiring grating couplers.
The integrated processor we utilize here even has built-in power monitors \cite{Morichetti.2014}, which can be seen in Fig.\,\ref{fig:ChipImage}\,(a), but they were not used for the presented experiments.

Fundamental to the photonic integrated processor is its interface to free space on the left (blue) which is actively used to sample free-space light distributions.
This interface can feature any layout of grating couplers, e.g., a regular grid-like layout similar to pixels of a normal camera.
However, our prototype device is restricted to 16 grating couplers and has thus been carefully designed to enable multiple applications.  
The chosen arrangement along two concentric rings with couplers oriented such that they are locally rotated by $\pm45^\circ$ with respect to the tangent line is, for example, well suited for applications requiring normal incidence or featuring rotational symmetry.

The final crucial aspect of the photonic integrated processor is how on-chip light is processed in the interferometric mesh which acts as a central light processing unit, see Fig.\,\ref{fig:ChipImage}\,(b).
Mach-Zehnder interferometers (yellow) arranged in a binary-tree can reroute the flow of light across the chip.
On-chip amplitudes and phases can be accurately controlled across the interferometric mesh. 
Light coupled into the circuit via the free-space interface and travelling through the mesh can, for example, be processed to be fully or partially combined in a single output waveguide thus enabling sorting, merging or adapting to free-space modes. 
Alternatively, by running light backward through the programmed mesh, this platform can also be used for other purposes. 
Modes of arbitrary relative intensity and phase could be generated losslessly in the circuit and emitted into free-space to form tailored distributions of structured light \cite{Butow.2023}, similar to optical phased arrays or beam shapers \cite{Sun.2013, Heck.2017}.

Before we describe how the photonic processor was used experimentally to measure higher-order beams we briefly discuss the utilization of the interferometric mesh.
Mentioned applications all rely on processing light in the central light processing unit while monitoring resulting waveguide intensities.
The straight forward approach is based on self-alignment and power minimizations \cite{Miller.2013b, Miller.2020}, while training the device with specifically designed input distributions of light.
A second approach relies on calibrating the mesh with a single reference input thereby characterizing all on-chip components simultaneously and utilizing this information to calculate the behavior of the mesh in subsequent applications.
Here we follow the latter approach, described in detail in Ref.\,\cite{Butow.2022}, and calibrate the mesh with a circularly polarized Gaussian beam of sufficient diameter (\SI{2}{\milli\meter}) to act as amplitude and phase reference.
With the calibrated mesh, on-chip amplitudes and phases of waveguide modes can be measured by monitoring the effect of the interferometric mesh on the transmitted intensities upon processing.
Unknown free-space distributions of light can thus be analyzed in regard to their relative amplitudes and phases at the grating coupler positions of the free-space interface. 

\begin{figure}[thb]
  \centering\includegraphics[width = 1\columnwidth]{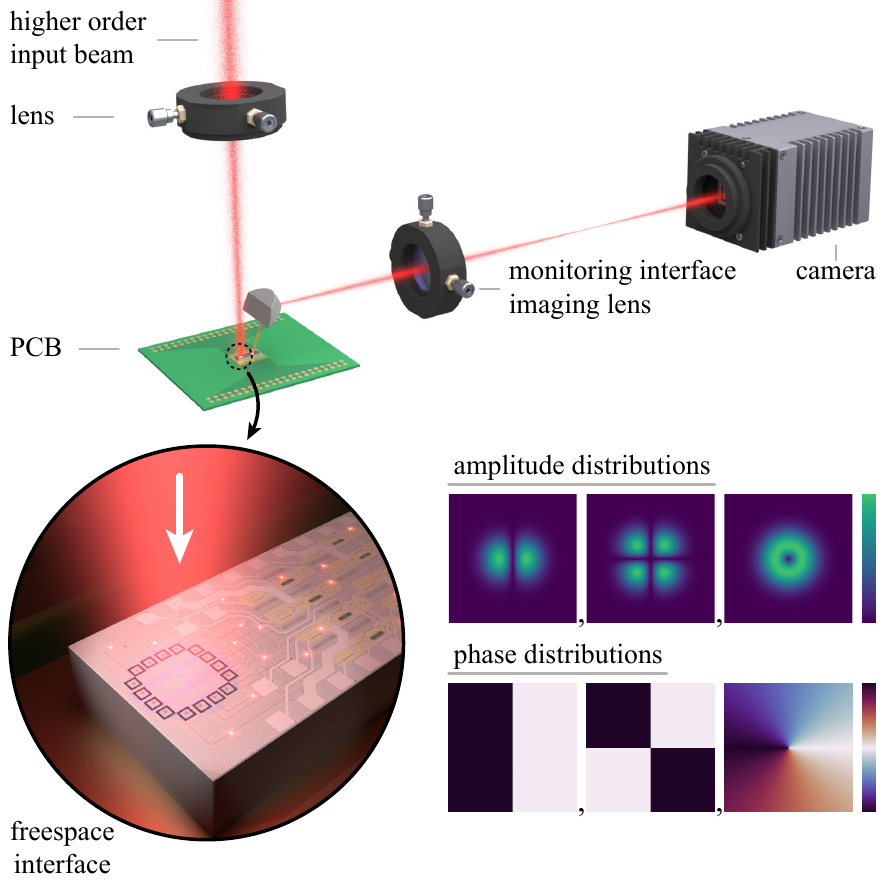}\caption{
    Illustration of the experimental setup to measure higher-order beams.
    These are weakly focused onto the photonic processor where they couple to on-chip waveguide modes which processed on-chip in a mesh of interferometers while transmitted intensities are recorded off-chip by imaging the monitoring interface.
    Ultimately, not only amplitude but also the phase information of the input beams is measured at each grating coupler position in the free-space interface.
}\label{fig:OpticalSetup}
\end{figure}

\vspace{.5cm}
\noindent
\textbf{Measuring higher-order beams.}
An illustration of the setups core components is shown in Fig.\,\ref{fig:OpticalSetup}.
A free-space laser beam with a wavelength of \SI{1570}{\nano\meter} is converted into a scalar spatial mode of light using a reflective phase-only spatial light modulator.
The resulting higher-order beam is polarized circularly before finally passing a lens of \SI{300}{\milli\meter} focal length.
The beam is weakly focused and thus matched in size to the photonic processor, which is wire-bonded to a printed circuit board (PCB).
The structured beam impinges normally onto the free-space interface where it couples to on-chip waveguide modes.
The resulting on-chip light travels across the interferometric mesh and is processed by reconfiguring all interferometers while the transmitted on-chip intensities are recorded.
In this experiment this is done off-chip by imaging the monitoring interface onto a conventional camera via a D-shaped pick-off mirror.

Now, we discuss what information the photonic processor can record in these measurements and what its advantages are compared to, e.g., a normal camera.
Nowadays, structured higher-order beams can be generated with ease and their properties with respect to intensity and phase distributions are well known theoretically.
Experimentally, their intensity information can be accessed by, for example, placing a camera in the beam path of, e.g., a Laguerre-Gaussian beam $\text{LG}_{0,1}$, which reveals its ring-like intensity distribution.
However, a camera is blind for the helical phase-front and the associated orbital angular momentum of such a beam.
In contrast, our photonic processor is capable of measuring amplitude and phase distributions simultaneously.
In case of the Laguerre-Gaussian beam this can be used to directly access the featured helical phase front and orbital angular momentum, as we will discuss in detail in the results section.

Note that while this information is not accessible to conventional cameras, other techniques and devices, mentioned in the introduction, are able to access the phase information of, e.g., higher-order beams. 
Our multipurpose photonic processor is a powerful and versatile addition to that list.

\section*{Results}
\begin{figure}[thb]
  \centering\includegraphics[width = 1\columnwidth]{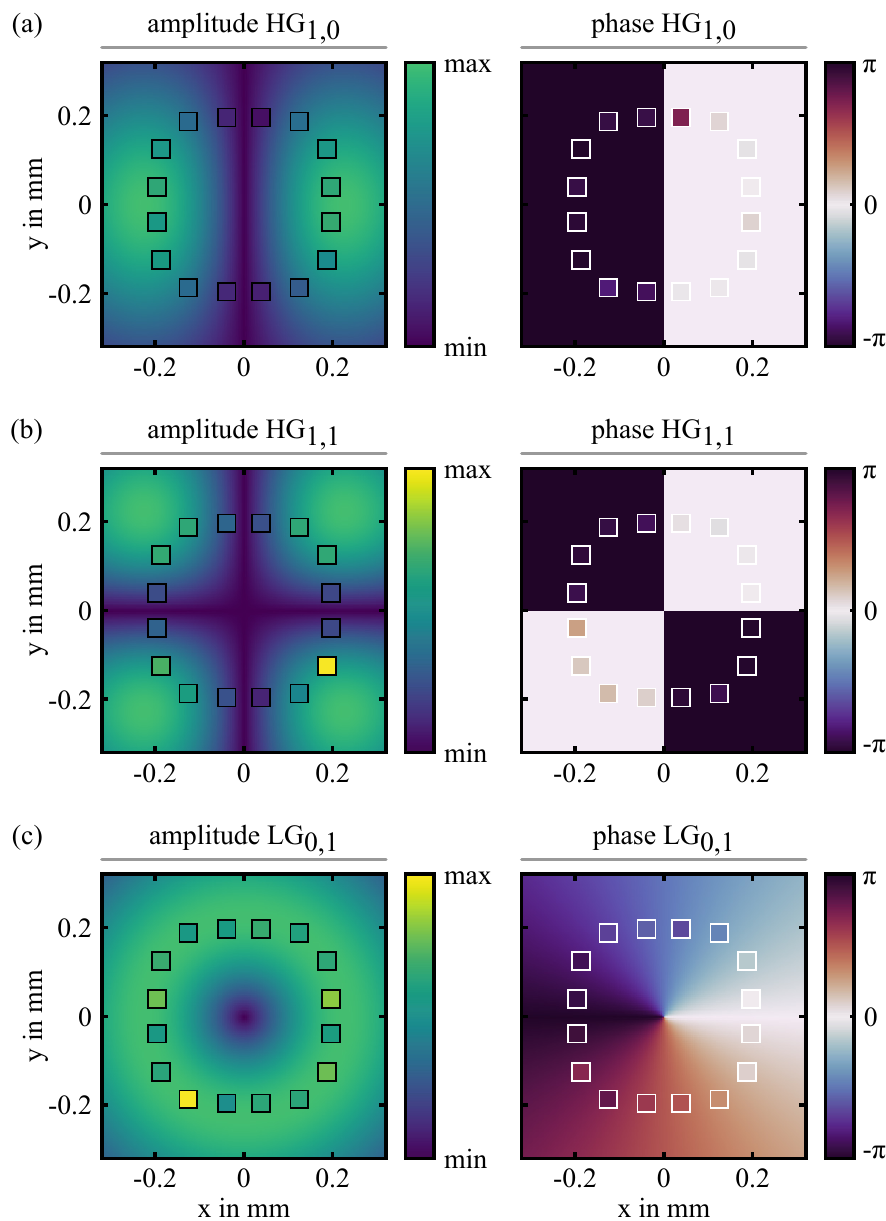}
  \caption{
    Amplitude and phase distributions of higher-order spatial modes. 
    Measured values are shown as squares at the 16 individual grating coupler positions.
    Theoretically calculated distributions are shown in the background.
    (a)~Hermite-Gaussian beam $\text{HG}_{1,0}$.
    (b)~Hermite-Gaussian beam $\text{HG}_{1,1}$.
    (c)~Laguerre-Gaussian beam $\text{LG}_{0,1}$.
  }\label{fig:ModesLGandHG}
\end{figure}

\noindent
\textbf{Phase and amplitude measurements.}
To demonstrate the capabilities of the integrated photonic processor in terms of measurements of higher-order beams and the identification of their orbital angular momentum, we measured different Hermite-Gaussian (HG) \cite{Svelto.2010} and Laguerre-Gaussian (LG) beams \cite{Allen.1992}.
In Fig.\,\ref{fig:ModesLGandHG} we show the theoretical as well as measured amplitude and phase profiles of a Hermite-Gaussian beam $\text{HG}_{1,0}$, a $\text{HG}_{1,1}$ beam and a Laguerre-Gaussian beam $\text{LG}_{0,1}$.
Theoretical distributions are plotted with high resolution in the background, and measured amplitude and phase values are superimposed as squares at the positions of the 16 individual grating couplers within the free-space interface of the photonic processor.
To allow for a qualitative comparison of relative amplitudes, we scaled our experimental data in Fig.\,\ref{fig:ModesLGandHG}\,(b) and (c) by a factor of $1.43$.
Theoretical expectation and experimental data are in very good agreement.
In detail, the measured phases nicely follow both the abrupt phase changes of the Hermite-Gaussian beams and the gradually azimuthally changing phase in case of the Laguerre-Gaussian beam.
For the latter, the phase clearly increases around the ring from 0 by $2\pi$, corresponding to an orbital angular momentum of 1, i.e., the azimuthal index of the $\text{LG}_{0,1}$ beam.
Minor deviations of the measured values could arise due to the intricate alignment of this few-pixels detector relative to the distributions of the light beam, modal imperfections of the input beams or minor systematic errors in the calibration, measurement or imaging of the photonic processor.
These results showcase how our photonic processor can measure higher-order beams and provide insights into their amplitude and phase information.

\begin{figure}[thb]
  \centering\includegraphics[width = 1\columnwidth]{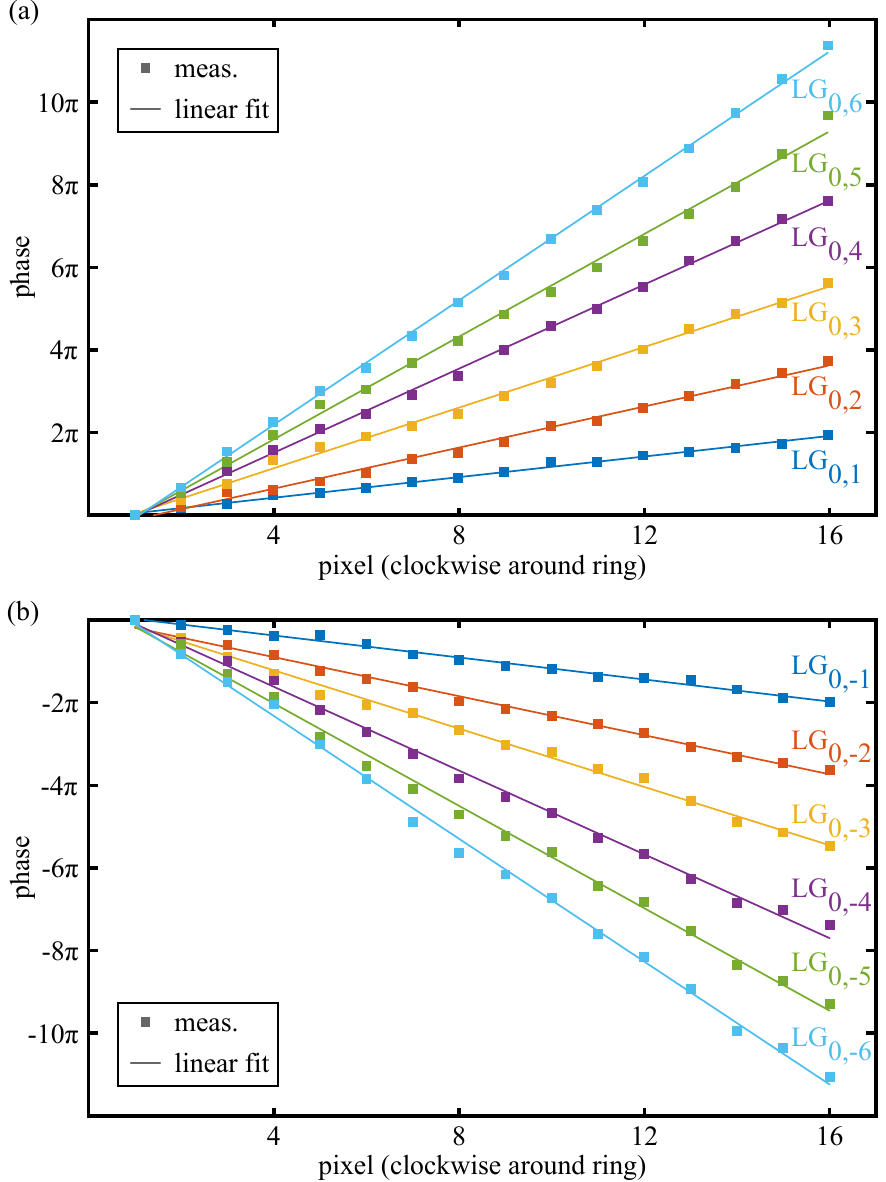}
  \caption{
    Measured phases (squares) of Laguerre-Gaussian beams of different azimuthal order 1\,to\,6 (a) and -1\,to\,-6 (b).
    The phase is measured at the grating coupler\,/\,pixel positions numbered clockwise around the free-space interface.
    The phase increases linearly (fits shown as solid lines) and the overall change in phase around the ring varies by multiples of $2\pi$.}
  \label{fig:OAMSorter}
\end{figure}

\vspace{.5cm}
\noindent
\textbf{Distinction and identification of orbital angular momentum.}
While the detector's resolution with 16 pixels is limited, it allows for distinguishing the orbital angular momentum of input beams.
The ring-like layout of the free-space interface does not resolve radial information with sufficient resolution, and we thus set the radial index of Laguerre-Gaussian input modes tested here to zero.
We only measure azimuthal changes along the ring-like layout and thus change the index $l$ of the input beams $\text{LG}_{0,l}$.
This index is associated with the beam's orbital angular momentum. 
For a given $\text{LG}_{0,l}$ beam, the relative phases between neighboring pixels in the transverse plane increase or decrease linearly, depending on the sense of the spiralling phase-front of the beam.
The overall phase change around the ring is $l\cdot2\pi$.
We show our experimental results of the measured phases for LG modes of azimuthal index 1~to~6 and -1~to~-6 in Fig.\,\ref{fig:OAMSorter}(a)~and~(b), respectively.
The measured phase values nicely follow the expected linear behavior with beams of higher azimuthal indices featuring a steeper slope.
By fitting a linear regression to the measured data we can extract the individual slopes $s$ of the curves and calculate the associated orbital angular momentum $l = 16s/2\pi$ of the various input beams.
We include these linear fits in Fig.\,\ref{fig:OAMSorter}.
The results of this orbital angular momentum retrieval method along with the resulting standard error are shown in Table\,\ref{tab:fittedOAM}.
\begin{table}[thb]
  \centering
  \caption{Measured orbital angular momentum and standard error of higher-order Laguerre-Gaussian input beams.}
  \begin{tabular}{ c c c c c }
   Beam & $\text{OAM}_\text{meas}$ & & Beam & $\text{OAM}_\text{meas}$ \\ 
   \hhline{==~==}
   $\text{LG}_{0,1}$ & 1.00 $\pm$ 0.02 & & $\text{LG}_{0,-1}$ & -1.07 $\pm$ 0.03 \\  
   $\text{LG}_{0,2}$ & 1.98 $\pm$ 0.04 & & $\text{LG}_{0,-2}$ & -1.89 $\pm$ 0.04 \\
   $\text{LG}_{0,3}$ & 2.93 $\pm$ 0.04 & & $\text{LG}_{0,-3}$ & -2.82 $\pm$ 0.05 \\  
   $\text{LG}_{0,4}$ & 4.06 $\pm$ 0.04 & & $\text{LG}_{0,-4}$ & -4.06 $\pm$ 0.06 \\
   $\text{LG}_{0,5}$ & 4.96 $\pm$ 0.07 & & $\text{LG}_{0,-5}$ & -4.96 $\pm$ 0.07 \\ 
   $\text{LG}_{0,6}$ & 6.01 $\pm$ 0.04 & & $\text{LG}_{0,-6}$ & -5.95 $\pm$ 0.08 \\
   \hhline{--~--}
  \end{tabular}
  \label{tab:fittedOAM}
\end{table}
The retrieved orbital angular momentum values match the azimuthal index of the Laguerre-Gaussian input beams very well, both in case of positive and negative azimuthal indices.
Deviations from the expected integer values of $l$ are small and could arise from modal impurities of the input beam or other small systematic measurement errors.
This shows that our photonic processor detects phases of higher-order beams very accurately and can be used to distinguish such beams based on phase information alone, which conventionally is difficult to access.

\section*{Discussion}
We have discussed and experimentally demonstrated the measurement of higher-order light beams using a multipurpose photonic integrated processor.
Its free-space interface in combination with the on-chip light processing resolves amplitudes at the grating coupler positions while simultaneously also locally measuring relative phases.
We showed the potential of such measurements by recording spatial amplitude and phase distributions of higher-order Hermite- and Laguerre-Gaussian beams and distinguishing Laguerre-Gaussian beams with azimuthal indices of up\,/\,down to $\pm 6$.
With only 16 pixels, this platform proves to be a powerful and versatile contender for characterizing orbital angular momentum and other free-space distributions of light.
Recent advances related to programmable photonic processors could be readily implemented, extending the presented multipurpose photonic processor platform and directly improving its ability to measure amplitude and phase distributions.
Larger and more generic input interfaces, polarization resolving grating couplers, the transition to visible wavelengths or full integration of the monitoring interface are just a small selection of possible improvements.
Already in its current form, the presented higher-order beam measurement of amplitude and phase as well as orbital angular momentum is a powerful tool and will find applications across various fields where, for instance, such knowledge gives deeper insights into the interaction processes light might have undergone with matter or the environment prior to detection.

\bibliographystyle{naturemag}
\footnotesize
\bibliography{modeMeasurementBibliography}
\normalsize

\section*{Materials and methods}
\noindent
\textbf{Silicon-on-insulator platform.} 
The reconfigurable photonic integrated circuit was fabricated commercially. 
Standard grating couplers, originally designed for fiber-coupling, where used to interface free-space light with on-chip waveguide modes. 
Each grating couples linearly polarized light to the transverse electric mode of the attached waveguide.
These single-mode waveguides have a width of \SI{500}{\nano\meter}. 
Each on-chip interferometer is composed of two directional couplers with a length of \SI{40}{\micro\meter} and a spacing of \SI{300}{\nano\meter}. 
To reconfigure each interferometer, two TiN heaters are placed above the waveguide, shifting the phase based on the thermo-optical effect.\\

\noindent
\textbf{Sampling free-space distributions.}
Our measurement approach is based on sampling impinging light beams with grating couplers arranged in a free-space interface and thus coupling light into waveguides, while preserving relative amplitudes and phases.
Each coupler is designed to couple linearly polarized light (polarized along the direction of the grating grooves) into the waveguides.
However, the local orientation of each grating coupler in the free-space interface is position dependent.
While this could be utilized to also measure the local polarization in the future, here we concentrate on amplitude and phase distributions only.
To avoid any dependences on polarization, we use circularly polarized input light.
Thus, each grating coupler locally sees and samples amplitude and phase distributions in its predefined polarization direction.
The azimuthally changing orientation of the grating couplers in the free-space interface combined with the circular input polarization results in an additional geometric phase, which we subtracted before displaying the data in this article.\\

\noindent
\textbf{Measurement procedure.} 
The data acquisition follows the same routine as described in Ref.\,\cite{Butow.2022}. 
We simultaneously scan the phase-shifters of all interferometers in the columns of the mesh by applying a grid of voltages, corresponding to a grid of 15-by-15 power values, to its two phase-shifters and recording the transmitted intensities. We repeat this for all columns of the mesh.
Voltage values between 0.2 and \SI{4}{\volt} where chosen to enable relative phase shifts between 0 and approximately $2\pi$.

\vspace{1cm}
\footnotesize
\noindent
\textbf{Acknowledgements} This work was supported by the European Commission through the H2020 project SuperPixels (grant 829116). The financial support by the Austrian Federal Ministry of Labour and Economy, the National Foundation for Research, Technology and Development and the Christian Doppler Research Association is gratefully acknowledged. The authors thank all members of the SuperPixels consortium for fruitful discussions and collaboration. We thank Maziyar Milanizadeh, Francesco Morichetti, Charalambos Klitis, and Marc Sorel for the photonic circuit design.

\noindent
\textbf{Disclosures} The authors declare no conflicts of interest.

\noindent
\textbf{Data availability} Data underlying the results presented in this paper are not publicly available at this time but may be obtained from the authors upon reasonable request.

\end{document}